\documentstyle[12pt]{article}
\begin{document}

\hspace*{50ex} OCU-PHYS-212 (Revised)  

\hspace*{50ex} September 2004 

\begin{center}
{\Large {\bf Ferromagnetism in quark matter} 
} \\

{\large {\sc A. Ni\'{e}gawa}\footnote{
E-mail: niegawa@sci.osaka-cu.ac.jp}

{\normalsize\em Graduate School of Science, Osaka City University } \\ 
{\normalsize\em Sumiyoshi-ku, Osaka 558-8585, Japan} } \\

\hspace*{2ex}

\hspace*{2ex}

\hspace*{2ex}

{\large {\bf Abstract}} \\ 
\end{center}

We investigate the magnetic property of the cold quark matter 
employing the magnetic moment $\vec{\mu}'$ as the order parameter. 
Through analysis of the effective potential $V (\vec{\mu}^{' 2})$, 
we find that, at relatively high densities $\rho > \rho_c$, the 
quark matter is in the normal phase. At $\rho = \rho_c$, the 
magnetic phase transition takes place and, in the low-density region 
$\rho < \rho_c$, the quark matter is in the ferromagnetic phase. For 
up, down and strange quarks, $\rho_c = 1.62 \rho_0$, $1.47 \rho_0$ 
and $2.20 \rho_0$ ($\rho_0$ the nuclear density), respectively. We 
also find that the leptons ($e$ and $\mu$) inside the quark matter 
are in the normal phase.

\newpage
%%%%%%%%%%%%%%%%%%%%%%%%%%%%%%%%%%%%%%%%%%%%%%%%%%%%%%%%%%%
\section{Introduction}
Several years ago, the possibility of existence of a quark matter in 
a ferromagnetic phase has been pointed out \cite{tatsumi}. 
Meanwhile, the possible discovery of a quark star \cite{1,2} has 
renewed our interest in this issue. 

In Ref. 1), through computing the energy density of a polarized 
quark matter, it is suggested that the Hartree-Fock state shows a 
spontaneous magnetic instability at low densities. The QCD coupling 
constant taken in Ref. 1) is $\alpha_s$ ($= g^2 / (4 \pi)$) $= 2.2$ 
and the quark mass is $m = 300$ MeV/c$^2$. In this paper, we examine 
a possibility of the quark matter in a ferromagnetic phase by 
choosing a magnetic moment $\vec{\mu}' \propto (e / m) \bar{\psi} 
\vec{\sigma} \psi$ as an order parameter. 

The QCD phase diagram is sketched in Fig. 1, where \lq $T$' is the 
temperature and \lq $\mu_B$' is the baryonic chemical potential. \lq 
QGP' stands for the quark-gluon plasma phase, \lq Hadron' stands for 
the hadronic phase and \lq CSC' stands for several phases which 
include the color super conducting phases. We are interested in the 
QGP phase near the phase boundaries and $0.6$ GeV $\leq \mu_B$. 
In this region $T << \mu_B$, so that we ignore the effect due to 
finite $T$. Thus, the object of our concern is the quark matter with 
finite baryon-number density at zero temperature. 

We have to introduce two kinds of cutoff and then the result depends 
on the ratio of these two. This is not particular to the present 
study but is a common feature in the study of effective potential 
for some order parameter (see, e.g., Ref. 4)). We shall simply 
equate the two cutoff parameters (Sec. 3), for which we take the 
baryonic chemical potential $\mu_B$ (Sec. 5). 

The plan of the paper is as follows. We introduce in Sec. 2 the 
closed-time-path (CTP) formalism \cite{le-b,sch,chou,lan} and derive 
the formula for the expectation value $\langle \vec{\mu}' \rangle$ 
under the existence of an uniform external magnetic field $\vec{B}$. 
In Sec. 3, within the leading logarithmic approximation, we compute 
$\langle \vec{\mu}' \rangle$ up to $O (\vec{B}^2 \vec{B})$ for 
$\vec{B}$ and to $O (g^2)$ for the QCD coupling constant. Then, 
performing the resummation of the leading contributions, we obtain 
the \lq\lq resummed form'' for $\langle \vec{\mu}' \rangle$. In Sec. 
4, we deal with the effective potential $V (\langle \vec{\mu}' 
\rangle)$. In Sec. 5, numerical analysis of $V (\langle \vec{\mu}' 
\rangle)$ is made in the range $0.6$ GeV $<\mu_B < 4$ GeV. We find 
that, in the high-density region $\mu_B > \mu_{Bc}$, the quark 
matter is in a normal phase. As the density is lowered, the quark 
matter undergoes a magnetic phase transition, at the critical 
density, into a ferromagnetic phase. The critical or transition 
chemical potentials are $\mu_{B c} = 641$, $625$ and $693$ MeV for 
up ($u$), down ($d$) and strange ($s$) quarks, respectively. We 
finally discuss the case of leptons and find that they are in the 
normal phase. Sec. 6 is devoted to summary and outlook. 
%%%%%%%%%%%%%%%%%%%%%%%%%%%%%%%%%%%%%%%%%%%%%%
\section{Preliminary}

Throughout this paper, we adopt the leading logarithmic 
approximation, 
%%%%%%%%%%%%%%%%%%%%%%%%%%%%%%%%%%%%%%%%%%%%%%%%%%%%%%%%%%%%%%%%
\begin{equation}
1 << \ln (\Lambda/m) \, , \;\;\;\;\;\;\;\;\; 1 << \ln [\mu_B / (3 
m)] \, . 
\label{app}
\end{equation}
%%%%%%%%%%%%%%%%%%%%%%%
Here, $\Lambda$ is the cutoff parameter and $\mu_B$ is the chemical 
potential being conjugate to the baryon number. The terms whose 
relative order of magnitude are $O(1 / \ln (\Lambda / m))$ and/or 
$O(1 / \ln [\mu_B / (3 m)])$ are ignored. 

As the order parameter, we adopt one-half of the spin magnetic 
moment, 
%%%%%%%%%%%%%%%%%%%
\[
\vec{\mu}' = \vec{\mu} / 2 = \frac{1}{2} \left( \frac{e_q}{2 m} 
\right) \bar{\psi} \vec{\sigma} \psi = \frac{e_q}{4 m} \bar{\psi} 
\gamma_5 \gamma_0 \vec{\gamma} \psi \, , 
\]
%%%%%%%%%%%%%%%%%%%%%
where and in the following the color index is suppressed. An 
argument for this choice is given in Appendix A. 
%%%%%%%%%%%%%%%%%%%%%%%%%%%%%%%%%%%%%%
%%%%%%%%%%%%%%  SUB SEC %%%%%%%%%%%%%%
%%%%%%%%%%%%%%%%%%%%%%%%%%%%%%%%%%%%%%
\subsection{Closed-time-path formalism} 
%%%%%%%%%%%%%%%%%%%%%%%%%%%%%%%%%%%%%%%
We use the closed-time-path (CTP) formalism 
\cite{le-b,sch,chou,lan}, which is formulated by introducing an 
oriented closed-time path $T_1 \otimes T_2$ in a complex time plane, 
that goes from $- \infty$ to $+ \infty$ ($T_1$) and then returned 
from $+ \infty$ to $- \infty$ ($T_2$). The field $\phi (x)$ with 
$x_0 \in T_i$ ($i = 1, 2$) is called type $i$, $\phi_i (x) = \phi 
(x) \, \rule[-2.5mm]{.14mm}{6.0mm} \raisebox{-2.3mm}{\scriptsize{$\; 
x_0 \in T_i$}}$. The Lagrangian density in this formalism is 
$\hat{\cal L}$ ($\equiv {\cal L} (\phi_1) - {\cal L} (\phi_2))$. 

The Lagrangian of a quark with some flavor reads 
%%%%%%%%%%%%%%%%%%%%%%%%%%%%%%%%%
\begin{eqnarray} 
\hat{\cal L} &=& \hat{\cal L}_{\mbox{\scriptsize{QCD}}} + 
\vec{\mu}_1' \cdot \vec{B}_1 - \vec{\mu}_2' \cdot \vec{B}_2 = 
\hat{\cal L}^{(2)} + \hat{\cal 
L}_{\mbox{\scriptsize{QCD}}}^{\mbox{\scriptsize{(int)}}} \, , 
\nonumber \\ 
\hat{\cal L}^{(2)} &=& \left( \bar{\psi}_1, \bar{\psi}_2 \right) 
\left[ \left( 
\begin{array}{cc}
1 & \; \, 0 \\ 
0 & \; \, - 1 
\end{array}
\right) 
\left( i 
\partial\mbox{\hspace*{0.3mm}}\kern-0.1em\raise0.3ex\llap{/}
\kern0.15em\relax - m \right) \right. \nonumber \\ 
&& \left. + \frac{e_q}{4 m} \gamma_5 \gamma_0 
\left( 
\begin{array}{cc}
\vec{\gamma} \cdot \vec{B}_1 & \; \, 0 \\ 
0 & \; \, - \vec{\gamma} \cdot \vec{B}_2 
\end{array}
\right) 
\right] \left( 
\begin{array}{c}
\psi_1 \\ 
\psi_2 
\end{array}
\right) 
\, , 
\label{bi}
\end{eqnarray} 
%%%%%%%%%%%%%%%%%%%%%%%%%%%%%%%%%%%%%%%%%%%%
where $\hat{\cal 
L}_{\mbox{\scriptsize{QCD}}}^{\mbox{\scriptsize{(int)}}}$ 
is the interaction part of $\hat{\cal L}_{\mbox{\scriptsize{QCD}}}$. 
It is convenient to construct a $(2 \times 2)$-matrix \lq\lq 
propagator'' $\hat{G}$, the inverse of the kernel of Eq. (\ref{bi}), 
which is diagonal in a color space. For the purpose of later use 
(cf. Eq. (\ref{kousiki})), it is sufficient to obtain $\hat{G}$ with 
$\vec{B}_1 = \vec{B}_2$ $(\equiv \vec{B})$. Straightforward 
manipulation yields for the $(i, j)$-element of $\hat{G}$, 
%%%%%%%%%%%%%%%%%%%%%%%%%%%%%%%%%%%%%%%%%%%%%%%%%%%
\begin{eqnarray} 
G_{11} (P) & = & - G_{22}^* (P) = G_F (P) - N (p_0) [G_F (P) - G_F^* 
(P)] \, , 
\label{11} \\ 
G_{12} (P) &=& [\theta (- p_0) - N (p_0)] [ G_F (P) - G_F^* 
(P)] \, , 
\label{12} \\ 
G_{21} (P) &=& [\theta (p_0) - N (p_0)] [ G_F (P) - G_F^* 
(P)] \, , 
\label{21} 
\end{eqnarray} 
%%%%%%%%%%%%%%%%%%%%%%%%%%%%%%%%%%%%%%%%%%%%%%%%%%%%%%%%%
where 
%%%%%%%%%%%%%%%%%%%%%%%%%%%%%%%%%%%%%%%%%%%%%%%%%%%
\begin{eqnarray} 
G_F (P) &=& \left[ \left\{ P^2 - m^2 + \left( \frac{e_q}{4 m} 
\right)^2 \vec{B}^2 \right\} \left( 
P\kern-0.1em\raise0.3ex\llap{/}\kern0.15em\relax + m + 
\frac{e_q}{4m} \gamma_5 \gamma_0 \vec{\gamma} \cdot \vec{B} \right) 
\right. \nonumber \\ 
&& - 2 \frac{e_q}{4 m} \vec{p} \cdot \vec{B} \left( - \frac{e_q}{4m} 
\vec{\gamma} \cdot \vec{B} + m \gamma_5 \gamma_0 - \gamma_5 
P\kern-0.1em\raise0.3ex\llap{/}\kern0.15em\relax \gamma_0 \right) 
\nonumber \\ 
&& \left. + 2 \frac{e_q}{4 m} p_0 \left( - \frac{e_q}{4m} \vec{B}^2 
\gamma_0 + m \gamma_5 \vec{\gamma} \cdot \vec{B} - \gamma_5 
P\kern-0.1em\raise0.3ex\llap{/}\kern0.15em\relax \vec{\gamma} \cdot 
\vec{B} \right) \right] \nonumber \\ 
&& \times \left[ \Delta_F^{- 2} (P) - 2 \left( \frac{e_q}{4 m} 
\right)^2 (\Delta_F^{- 1} (P) + 2 E_p^2) \vec{B}^2 \right. 
\nonumber \\ 
&& \left. + 4 \left( \frac{e_q}{4 m} \right)^2 (\vec{p} \cdot 
\vec{B} )^2 + \left( \frac{e_q}{4 m} \right)^4 (\vec{B}^2)^2 
\right]^{- 1} \, , 
\label{takara} 
\\ 
\Delta_F (P) &=& \frac{1}{P^2 - m^2 + i 0^+} \nonumber \, . 
\end{eqnarray} 
%%%%%%%%%%%%%%%%%%%%%%%%%%%%%%%%%%%%%%%%%%%%%%%%%%%%%%%%%
Here we have used a capital letter for denoting a four vector $P = 
(p_0, \vec{p})$ $(p \equiv |\vec{p}|)$ and $E_p = \sqrt{p^2 + m^2}$. 
In the above equations, \lq\lq $\,{}^*\,$'' does not apply to the 
Dirac gamma matrices and $N (p^0)$ is the number density function, 
%%%%%%%%%%%%%%%%%%%%%%%%%%%%%%%%
\begin{equation} 
N (p_0) = \theta (p_0) \theta (\mu_B / 3 - p_0) \, . 
\label{bunpu} 
\end{equation} 
%%%%%%%%%%%%%%%%%%%%%%%%%%%%%%%%%%%%%%%%%%%%%%%%%%%

\hspace*{3ex}

\begin{center} 
{\em Generating functional and the formula for $\langle \vec{\mu}' 
\rangle$} 
\end{center}

\hspace*{3ex}

The CTP generating functional is defined \cite{chou} by 
%%%%%%%%%%%%%%%%%%%
\[
Z (\vec{B}_1, \vec{B}_2) = \int \prod_{j = 1}^2 \left[ {\cal D} 
\psi_j {\cal D} \bar{\psi}_j {\cal D} A_j \right] \exp \left[i \int 
d^{\, 4} x \left( \hat{\cal L}_{\mbox{\scriptsize{QCD}}} + 
\vec{\mu}_1' \cdot {\vec B}_1 - \vec{\mu}_2' \cdot {\vec B}_2 
\right) \right] \rho \, , 
\]
%%%%%%%%%%%%%%%%%%%%%%
where $A_j$ ($j = 1, 2$) is the gluon field and $\rho$ is the 
density matrix. An expectation value of the magnetic moment 
$\vec{\mu}'$ is computed through \cite{chou} 
%%%%%%%%%%%%%%%%%%%
\begin{eqnarray}
\langle \vec{\mu}' \rangle &\equiv& \frac{1}{2} \left( \langle 
\vec{\mu}_1' \rangle + \langle \vec{\mu}_2' \rangle \right) 
\, \rule[-3mm]{.14mm}{8.5mm} \raisebox{-2.85mm}{\scriptsize{$\, 
\vec{B}_1 = \vec{B}_2 \equiv \vec{B}$}} 
\nonumber \\ 
&=& - \frac{i}{2} \left( \frac{\delta}{\delta \vec{B}_1} - 
\frac{\delta}{\delta \vec{B}_2} \right) \ln Z (\vec{B}_1, \vec{B}_2) 
\, \rule[-3mm]{.14mm}{8.5mm} \raisebox{-2.85mm}{\scriptsize{$\, 
\vec{B}_1 = \vec{B}_2 \equiv \vec{B}$}} \, . 
\label{kousiki}
\end{eqnarray}
%%%%%%%%%%%%%%%%%%%%%%
We compute $\langle \vec{\mu}' \rangle$ up to $O (\vec{B}^2 
\vec{B})$. 
%%%%%%%%%%%%%%%%%%%%%%%%%%%%%%%%%%%%%%%%%%%%%%%%%%%%
%% SEC %%%%%%%%%%%%%%%%%%%%%%%%%%%%%%%%%%%%%%%%%
%%%%%%%%%%%%%%%%%%%%%%%%%%%%%%%%%%%%%%%%%%%%%%%%
\section{Computation of 
$\langle \vec{\mu}' \rangle$}
\subsection{0th order contribution}
%%%%%%%%%%%%%%%%%%%
From Eq. (\ref{kousiki}), we obtain for the $0$th-order contribution 
(${\cal L}_{\mbox{\scriptsize{QCD}}}^{\mbox{\scriptsize{(int)}}} = 
0$) to $\langle \vec{\mu}' \rangle$, 
%%%%%%%%%%%%%%%%%%%
\begin{eqnarray}
\langle \vec{\mu}' \rangle & = & - \frac{i}{2} \frac{e_q}{4 m} 
\mbox{Tr} \int 
\frac{d^{\, 4} P}{(2 \pi)^4} \left[ \gamma_5 \gamma_0 \vec{\gamma} 
\left( G_{11} (P) + G_{22} (P) \right) \right] \nonumber \\ 
&=& \frac{e_q}{4 m} Im \mbox{Tr} \int \frac{d^{\, 4} P}{(2 \pi)^4} 
\left[ \gamma_5 \gamma_0 \vec{\gamma} G_{11} (P) \right] 
\, . 
\label{zero} 
\end{eqnarray}
%%%%%%%%%%%%%%%%%%%%%%
Straightforward manipulation using Eq. (\ref{takara}) yields, up to 
$O (\vec{B}^2 \vec{B})$, 
%%%%%%%%%%%%%%%%%%%%%%%%%%%%%%%%%%%%%%%%%%%%%%%%%%%
\begin{eqnarray} 
&& Im \mbox{Tr} \int \frac{d^{\, 4} P}{(2 \pi)^4} \left[ \gamma_5 
\gamma^0 \vec{\gamma} G_F (P) \right] \nonumber \\ 
&& \mbox{\hspace*{5ex}} = - \frac{4}{3} \frac{e_q \vec{B}}{4 m} Im 
\int \frac{d^{\, 4} P}{(2 \pi)^4} \left[ 3 \Delta_F + 2 (2 p^2 + 3 
m^2) \Delta_F^2 \right] \nonumber \\ 
&& \mbox{\hspace*{7.7ex}} - \frac{4}{15} \left( \frac{e_q}{4 m} 
\right)^3 \vec{B}^2 \vec{B} \, Im \int \frac{d^{\, 4} P}{(2 \pi)^4} 
\left[ 15 \Delta_F^2 \right. \nonumber \\ 
&& \mbox{\hspace*{7.7ex}} \left. + 40 (2 p^2 + 3 m^2) \Delta_F^3 + 8 
(8 p^4 + 20 m^2 p^2 + 15 m^4) \Delta_F^4) \right] \, . 
\label{zero1}
\end{eqnarray} 
%%%%%%%%%%%%%%%%%%%%%%%%%%%%%%%%%%%%%%%%%%%%%%%%%%%%%%%%%%
Substituting Eq. (\ref{11}) for $G_{11}$ in Eq. (\ref{zero}) and 
using Eq. (\ref{zero1}), one can compute $\langle \vec{\mu}' 
\rangle$. 

We divide $\langle \vec{\mu}' \rangle$ into two pieces, 
%%%%%%%%%%%%%%%%%%%%%%%%%%%%%%%
\[
\langle \vec{\mu}' \rangle = \langle \vec{\mu}' 
\rangle_{\mbox{\scriptsize{vac}}} + \langle \vec{\mu}' 
\rangle_{\mbox{\scriptsize{mat}}} \, , 
\]
%%%%%%%%%%%%%%%%%%%%%%
where $\langle \vec{\mu}' \rangle_{\mbox{\scriptsize{vac}}}$ 
($\langle \vec{\mu}' \rangle_{\mbox{\scriptsize{mat}}}$) stands for 
the contribution from the vacuum (matter) sector. $\langle 
\vec{\mu}' \rangle_{\mbox{\scriptsize{vac}}}$ is given by Eq. 
(\ref{zero}) with $G_F$ for $G_{11}$ (cf Eq. (\ref{11})). Thanks to 
the $O (3, 1)$ symmetry, in Eq. (\ref{zero1}), the replacements, 
$p^2 \to - 3 P^2 / 4$ and $p^4 \to - 5 (P^2)^2 / 8$ may be made. 
Thus we obtain 
%%%%%%%%%%%%%%%%%%%%%%%%%%%%%%%%%%%%
\begin{equation} 
\langle \vec{\mu}' \rangle_{\mbox{\scriptsize{vac}}} = - 4 \left( 
\frac{e_q}{4 m} \right)^2 \vec{B} \, Im \int \frac{d^{\, 4} P}{(2 
\pi)^4} \, m^2 \Delta_F^2 (P) 
+ \frac{4}{3} \left( \frac{e_q}{4 m} \right)^4 \vec{B}^2 \vec{B} 
\, Im \int \frac{d^{\, 4} P}{(2 \pi)^4} \Delta_F^2 (P) \, . 
\label{mei} 
\end{equation} 
%%%%%%%%%%%%%%%%%%
In the second term on the right-hand side (RHS), we have dropped the 
terms that are proportional to $m^2 (e_q / 4 m)^4$ and $m^4 (e_q / 
4 m)^4$, the contribution of which is nonleading in our 
approximation (cf. Eq. (\ref{app})). Introducing a momentum cutoff 
$\Lambda$, we have 
%%%%%%%%%%%%%%%%%%%%%%%%%
\begin{equation} 
\int \frac{d^{\, 4} P}{(2 \pi)^4} \, \Delta_F^2 (P) = \frac{i}{8 
\pi^2} \ln (\Lambda / m) \, , 
\label{cut} 
\end{equation} 
%%%%%%%%%%%%%%%%%%%%%%%
with which we obtain the first terms on the RHS's of Eqs. 
(\ref{vs1}) and (\ref{vs2}) below. 

For $\langle \vec{\mu}' \rangle_{\mbox{\scriptsize{mat}}}$, we 
obtain 
%%%%%%%%%%%%%%%%%%%%%%%%%%%%%%%%%%%%
\begin{eqnarray} 
\langle \vec{\mu}' \rangle_{\mbox{\scriptsize{mat}}} &=& \frac{8}{3} 
\left( \frac{e_q}{4 m} \right)^2 \vec{B} \, Im \int \frac{d^{\, 4} 
P}{(2 \pi)^4} \theta (\mu_B / 3 - p_0) \nonumber \\ 
&& \times \left[ 3 \Delta_F + 4 p^2 \Delta_F^2 
%\right. \nonumber \\ 
%&& \left. 
+ \frac{1}{5} \left( \frac{e_q \vec{B}}{4 m} \right)^2 
\left( 15 \Delta_F^2 + 80 p^2 \Delta_F^3 + 64 p^4 \Delta_F^4 
\right) \right] \, . \nonumber \\ 
&& 
\label{meic} 
\end{eqnarray} 
%%%%%%%%%%%%%%%%%%
Here the terms that are proportional to $m^2 (e_q / 4 m)^{2 n}$ and 
$m^4 (e_q / 4 m)^{2 n}$ $(n = 1, 2)$ have been dropped, which yield 
the subleading contributions. 
%%%%%%%%%%%%%%%%%%%%%%
\subsection{$O(g^2)$ QCD correction to $\langle \vec{\mu}' \rangle$} 
%%%%%%%%%%%%%%%%%%%
Here we compute the $O(g^2)$ QCD correction to $\langle \vec{\mu}' 
\rangle$ in Eq. (\ref{kousiki}). Gluon propagator is diagonal in a 
color space. Choosing a covariant 
gauge, we have 
%%%%%%%%%%%%%%%%%%%%%%%%%%%%%%%%%%%%
\begin{eqnarray} 
D_{11}^{\nu \mu} (Q) & = & - \left( D_{22}^{\nu \mu} (Q) \right)^* 
= D_F^{\nu \mu} (Q) = - \frac{g^{\nu \mu}}{Q^2 + i 0^+} + \eta 
\frac{Q^\nu Q^\mu}{(Q^2 + i 0^+)^2} \, , 
\label{p11} \\ 
D_{12}^{\nu \mu} (Q) &=& \theta (- q^0) [ D_F^{\nu \mu} (Q) - \left( 
D_F^{\nu \mu} (Q)\right)^* ] \, , 
\label{UK} \\ 
D_{21}^{\nu \mu} (Q) &=& \theta (q^0) [ [ D_F^{\nu \mu} (Q) - 
\left(D_F^{\nu \mu} (Q)\right)^* ] \, . 
\label{D} 
\end{eqnarray} 
%%%%%%%%%%%%%%%%%%
To $O (g^2)$, we obtain from Eq. (\ref{kousiki}), 
%%%%%%%%%%%%%%%%%%%
\begin{eqnarray} 
\langle \vec{\mu}' \rangle &=& - \frac{i}{2} \frac{e_q}{4 m} 
\sum_{j, l = 1}^2 \mbox{Tr} \int \frac{d^{\, 4} P}{(2 \pi)^4} 
\gamma_5 \gamma_0 \vec{\gamma} 
\left[ \left\{ G_{1 j} (P) \tilde{\Sigma}_{l j} (P) 
G_{l 1} (P) \right\} + \left\{ 1 \to 2 \right\} \right] \, , 
\nonumber \\ 
\tilde{\Sigma}_{j l} (P) & = & \frac{4 i}{3} g^2 (-)^{j + l} \int 
\frac{d^{\, 4} Q}{(2 \pi)^4} \gamma^\mu G_{j l} (P + Q) 
\gamma^\nu D_{l j}^{\nu \mu} (Q) \nonumber \, . 
\end{eqnarray} 
%%%%%%%%%%%%%%%%%%%%%%%%%%%%%%%%%%%%%%%%%%%%%%%%%%%%
Using Eqs. (\ref{11}) - (\ref{21}) and (\ref{p11}) - (\ref{D}), we 
get, after some manipulation, 
%%%%%%%%%%%%%%%%%%%
\begin{eqnarray} 
\langle \vec{\mu}' \rangle &=& \frac{e_q}{4 m} Im \mbox{Tr} \int 
\frac{d^{\, 4} P}{(2 \pi)^4} \gamma_5 \gamma_0 \vec{\gamma} 
[1 - 2 N (p_0)] G_F (P) \tilde{\Sigma}_F (P) G_F (P) 
\, , 
\label{mualp} \\ 
\tilde{\Sigma}_F (P) &=& \tilde{\Sigma}_{11} (P) + \theta (p_0) 
\tilde{\Sigma}_{12} (P) + \theta (- p_0) \tilde{\Sigma}_{21} (P) 
\nonumber \\ 
&=& \frac{4 i}{3} g^2 \int \frac{d^{\, 4} Q}{(2 \pi)^4} \gamma^\mu 
\left[ G_{11} (P + Q) D_F^{\nu \mu} (Q) - \theta (p_0) G_{12} (P + 
Q) D_{21}^{\nu \mu} (Q) \right. \nonumber \\ 
&& \left.- \theta (- p_0) G_{21} (P + Q) D_{12}^{\nu \mu} (Q) 
\right] \gamma^\nu \nonumber \\ 
&=& \frac{4 i}{3} g^2 \int \frac{d^{\, 4} Q}{(2 \pi)^4} \gamma^\mu 
\left[ G_{11} (P + Q) D_F^{\nu \mu} (Q) \right. \nonumber \\ 
&& + N (p_0 + q_0) \left\{ G_F (P + Q) - G_F^* (P + Q) 
\right\} \nonumber \\ 
&& \left. \times \left\{ \theta (p_0) D_{21}^{\nu \mu} (Q) + \theta 
(- p_0) D_{12}^{\nu \mu} (Q) \right\} \right] \gamma^\nu \nonumber 
\, . 
\end{eqnarray} 
%%%%%%%%%%%%%%%%%%%%%%%%%%%%%%%%%%%%%%%%%%%%%%%%%%%%
The term with $\theta (- p^0) D_{12}^{\nu \mu} (Q)$ in the last line 
vanishes, since $D_{12}^{\nu \mu} (Q) = 0$ for $q_0 > 0$, $N (p_0 + 
q_0) = 0$ for $p_0 + q_0 < 0$. Nevertheless we include it for later 
convenience. Using Eqs. (\ref{11}), (\ref{UK}), and (\ref{D}), we 
obtain 
%%%%%%%%%%%%%%%%%%%%%%%%%%%%%%
\begin{eqnarray}
\tilde{\Sigma}_F (P) &=& \frac{4 i}{3} g^2 \int \frac{d^{\, 4} Q}{(2 
\pi)^4} \gamma^\mu \left[ G_F (P + Q) D_F^{\nu \mu} (Q) \right. 
\nonumber \\ 
&& \left. - N (p_0 + q_0) \left\{ G_F (P + Q) - G_F^* (P + Q) 
\right\} \right. \nonumber \\ 
&& \times \left\{ 
\theta (p_0) \left( 
\theta (- q_0) D_F^{\nu \mu} (Q) + 
\theta (q_0) \left( D_F^{\nu \mu} (Q) \right)^* 
\right) \right. \nonumber \\ 
&& \left. \left. + \theta (- p_0) \left( 
\theta (q_0) D_F^{\nu \mu} (Q) + \theta (- q_0) \left( D_F^{\nu \mu} 
(Q) 
\right)^* \right) \right\} \right] \gamma^\nu \nonumber \, . 
\end{eqnarray}
%%%%%%%%%%%%%%%%%%%%%%%%%%%%%%%%
Substituting this for $\tilde{\Sigma}_F (P)$ in Eq. (\ref{mualp}), 
we finally obtain 
%%%%%%%%%%%%%%%%%%%
\begin{eqnarray} 
\langle \vec{\mu}' \rangle &=& \langle \vec{\mu}' \rangle_1 + 
\langle \vec{\mu}' \rangle_2 \, , 
\label{bunkai} \\ 
\langle \vec{\mu}' \rangle_1 &=& \frac{4}{3} g^2 \frac{e_q}{4 m} Re 
\mbox{Tr} \int \frac{d^{\, 4} P}{(2 \pi)^4} \int \frac{d^{\, 4} 
Q}{(2 \pi)^4} \gamma_5 \gamma_0 \vec{\gamma} \nonumber \\ 
&& \times [1 - 2 N (p_0)] G_F (P) \gamma^\mu G_F (P + Q) D_F^{\nu 
\mu} (Q) \gamma^\nu G_F (P) \, , 
\label{3-1} \\ 
\langle \vec{\mu}' \rangle_2 &=& - \frac{4}{3} g^2 \frac{e_q}{4 m} 
Re \mbox{Tr} \int \frac{d^{\, 4} P}{(2 \pi)^4} \int \frac{d^{\, 4} 
Q}{(2 \pi)^4} \gamma_5 \gamma_0 \vec{\gamma} \nonumber \\ 
&& \times [1 - 2 N (p_0)] G_F (P) \gamma^\mu N (p_0 + q_0) 
\left\{ G_F (P + Q) - G_F^* (P + Q) \right\} \nonumber \\ 
&& \times \left\{ \theta (p_0) D_A^{\nu \mu} (Q) + \theta (- p_0) 
D_R^{\nu \mu} (Q) \right\} \gamma^\nu G_F (P) \, , 
\label{3-2} 
\end{eqnarray} 
%%%%%%%%%%%%%%%%%%%%%%%%%%%%%%%%%%%%%%%%%%%%%%%%%%%%
where $D_R^{\nu \mu}$ ($D_R^{\nu \mu}$) is the retarded (advanced) 
gluon propagator; 
%%%%%%%%%%%%%%%%%%%%%%%%%%%%%%%%%%%%%%%%%%%%%%%%%%%%%%%%%%%%
\[ 
D_R (Q) = [D_A (Q)]^* = \theta (q_0) D_F (Q) + \theta (- q_0) 
[D_F (Q)]^* \, . 
\]
%%%%%%%%%%%%%%%%%%%%%%%%%%%%
The vacuum-sector contribution is given by Eq. (\ref{3-1}) with 
$N (p_0) = 0$. 

It is necessary to compute (up to $O (\vec{B}^2 \vec{B})$) the 
quantity, 
%%%%%%%%%%%%%%%%%%%%%%%%
\[ 
\vec{\cal G} \equiv \mbox{Tr} \left[ \gamma_5 \gamma_0 
\vec{\gamma} G_F (P) \gamma^\mu G_F (P') \gamma_\mu G_F (P) \right] 
D_T (Q) \;\;\;\;\;\;\;\; (P' = P + Q) \, , 
\] 
%%%%%%%%%%%%%%%%%%%%%%%%%
where $T$ stands for $F, R$, or $A$. Introducing 
%%%%%%%%%%%%%%%%%%%%%%%%
\[ 
{\cal I}_{lmn}^{(T)} \equiv [\Delta_F (P)]^l [\Delta_F (P')]^m 
\left( [\Delta_T (Q)]_{m = 0} \right)^n \;\;\;\;\;\;\;\; (T = F, R, 
A) \, , 
\] 
%%%%%%%%%%%%%%%%%%%%%%%%%
we obtain using Eq. (\ref{takara}), 
%%%%%%%%%%%%%%%%%%%
\begin{eqnarray} 
\vec{\cal G} &=& \frac{e_q}{4 m} \vec{B} \left( {\cal 
G}_{\mbox{\scriptsize{self}}}^{(1)} + {\cal G}^{(1)'} \right) + 
\left( \frac{e_q}{4 m} \right)^3 \vec{B}^2 \vec{B} \left( {\cal 
G}_{\mbox{\scriptsize{self}}}^{(2)} + {\cal G}^{(2)'} \right) \, , 
\label{abc} \\ 
{\cal G}_{\mbox{\scriptsize{self}}}^{(1)} &=& - 8 \left[ {\cal I}_{1 
1 1}^{(T)} + \left\{ 3 (Q^2 + 2 m^2) - \frac{4}{3} \vec{p} \cdot 
\vec{q} + 4 p_0 q_0 \right\} {\cal I}_{2 1 1}^{(T)} \right. 
\nonumber \\ 
&& \left. + \frac{8}{3} 
p^2 \left( Q^2 + 2 m^2 \right) {\cal I}_{3 
1 1}^{(T)} \right] \, , \\ 
{\cal G}^{(1)'} &=& - \frac{32}{3} m^2 \left[ 3 p_0 p_0' - 
\vec{p} \cdot \vec{p'} \right] {\cal I}_{221}^{(T)} \, , \\ 
{\cal G}_{\mbox{\scriptsize{self}}}^{(2)} &=& - 16 \left[ 11 {\cal 
I}_{2 1 1}^{(T)} + \left( - 11 Q^2 + 8 p^2 + 8 p_0 p_0' - 
\frac{8}{3} \vec{p} \cdot \vec{p'} \right) {\cal I}_{3 
1 1}^{(T)} \right. \nonumber \\ 
&& \left. + 8 p^2 \left( - Q^2 + \frac{4}{3} p_0 p_0' - \frac{4}{15} 
\vec{p} \cdot \vec{p'} \right) {\cal I}_{4 1 1}^{(T)} \right] \, , 
\label{abd} \\ 
{\cal G}^{(2)'} &=& - 8 \left[ - 9 \left( {\cal I}_{211}^{(T)} + 
{\cal I}_{121}^{(T)} - {\cal I}_{220}^{(T)} \right) + 8 \left( 2 p_0 
p_0' - p^{' 2} - p^2 - \frac{2}{3} \vec{p} \cdot \vec{p'} \right) 
{\cal I}_{221}^{(T)} \right. \nonumber \\ 
&& + 8 p^{' 2} {\cal I}_{230}^{(T)} + \frac{32}{15} \left( 5 p_0 
p_0' p^{' 2} - 3 p^2 p^{' 2} - p^{' 2} \vec{p} \cdot \vec{p'} - 
(\vec{p} \cdot \vec{p'})^2 \right) {\cal I}_{231}^{(T)} \nonumber \\ 
&& - 8 p^{' 2} {\cal I}_{131}^{(T)} - 8 p^2 {\cal I}_{311}^{(T)} + 
\frac{32}{15} \left( 3 p^2 p^{' 2} + (\vec{p} \cdot \vec{p'})^2 
\right) {\cal I}_{330}^{(T)} - \frac{8}{3} p^2 {\cal I}_{320}^{(T)} 
\nonumber \\ 
&& \left. + \frac{32}{3} p^2 {\cal I}_{320}^{(T)} + \frac{32}{15} 
\left( - 3 p^2 p^{' 2} - p^2 \vec{p} \cdot \vec{p'} + 5 p_0 p_0' p^2 
- (\vec{p} \cdot \vec{p'})^2 \right) {\cal I}_{321}^{(T)} \right] 
\, . \nonumber \\ 
&& 
\label{xyz}
\end{eqnarray} 
%%%%%%%%%%%%%%%%%%%%%%%%%%%%%%%%%%%%%%%%%%%%%%%%%%%%
Here ${\cal G}_{\mbox{\scriptsize{self}}}^{(j = 1, 2)}$ comes from 
the diagram that includes a quark self-energy subdiagram. For ${\cal 
G}_{\mbox{\scriptsize{self}}}^{(2)} + {\cal G}^{(2) '}$, the 
contributions that are proportional to $m^{2n}$ ($n \geq 1$) has 
been dropped, which do not yield leading contributions. 

According to Eqs. (\ref{abc}) - (\ref{xyz}), we write $\langle 
\vec{\mu}' \rangle_j$ ($j = 1, 2$) in Eq. (\ref{bunkai}), with 
obvious notation, as 
%%%%%%%%%%%%%%%%%%%%%%%%%%%%%%%
\begin{equation} 
\langle \vec{\mu}' \rangle_j = \langle \vec{\mu}' 
\rangle_j^{\mbox{\scriptsize{(self)}}} + \langle \vec{\mu}' 
\rangle_j' \;\;\;\;\;\;\;\;\; (j = 1, 2) \, . 
\label{mei1} 
\end{equation} 
%%%%%%%%%%%%%%%%%%%%%%%%%%%%%%%%%%%%%%%%%%%%%%%%%%%%%
The quark self-energy part $\Sigma (P)$ consists of two pieces, 
$\Sigma (P) = \Sigma_{\mbox{\scriptsize{vac}}} (P) + 
\Sigma_{\mbox{\scriptsize{mat}}} (P)$. Here 
$\Sigma_{\mbox{\scriptsize{vac}}} (P)$ stands for the self-energy 
part in vacuum theory, while $\Sigma_{\mbox{\scriptsize{mat}}} (P)$ 
is due to the existence of matter. Eqs. (\ref{3-1}) and (\ref{3-2}) 
tell us that $\langle \vec{\mu}' 
\rangle_1^{\mbox{\scriptsize{(self)}}}$ includes 
$\Sigma_{\mbox{\scriptsize{vac}}} (P)$ and $\langle \vec{\mu}' 
\rangle_2^{\mbox{\scriptsize{(self)}}}$ includes 
$\Sigma_{\mbox{\scriptsize{mat}}} (P)$. 
%%%%%%%%%%%%%%%%%%%%%%%%%%%%%%%%%%%%%%%%%%%%%%%%%%%%%%%%%%
\subsubsection*{Observation on Eqs. (\ref{3-1}) and (\ref{3-2})} 
%%%%%%%%%%%%%%%%%%%%%%%%%%%%%%%%%%%%%%%%%%%%%%%%%%%%%%%%%%
It is not difficult to see that, both in Eq. (\ref{3-1}) and 
in (\ref{3-2}), an integration over $P$ (with $Q$ fixed) is free 
from ultraviolet divergence. Then, we introduce a Pauli-Villars 
regulator (with mass $\Lambda_{\mbox{\scriptsize{PV}}}$) for a 
gluon propagator, which makes the integrals in Eq. (\ref{3-1}) and 
in Eq. (\ref{3-2}) converge. This means that the contribution from 
the gauge-parameter $(\eta)$ dependent part of the gluon propagator 
(Eq. (\ref{p11})) vanishes, thanks to the Ward identity, and the 
results (\ref{vs1}) - (\ref{end}) below are independent of 
$\eta$. Thus, we use the Feynman gauge, $\eta = 0$. In the following 
we equate $\Lambda_{\mbox{\scriptsize{PV}}}$ with $\Lambda$ in Eq. 
(\ref{cut}). 
%%%%%%%%%%%%%%%%%%%%%%%%%%%%%%%%%%%%%%%%%%%%%%%%%%%%%%%%%%
\subsection{Form for $\langle \vec{\mu}' \rangle$} 
Here we display the resultant form for $\langle \vec{\mu}' \rangle$, 
%%%%%%%%%%%%%%%%%%%%%%%%%%%%%%%%%%%
\begin{equation} 
\langle \vec{\mu}' \rangle = \left[ C_1 + \vec{B}^2 C_2 \right] 
\vec{B} \, . 
\label{mu}
\end{equation} 
%%%%%%%%%%%%%%%%%%%%%%%%%%%%%%
Summing over colors, we have for the contributions to $C_1$ and to 
$C_2$ from the vacuum sector and those owing to the existence of 
matter, in respective order, 
%%%%%%%%%%%%%%%%%%%%%%%%%%%%%%
\begin{eqnarray}
C_1^{\mbox{\scriptsize{(vac)}}} &=& - \frac{3 e^2_q}{32 \pi^2} 
\left[ 1 + \frac{4 \alpha_s}{3 \pi} \ln \frac{\Lambda}{m} \right] 
\ln \frac{\Lambda}{m} \, , 
\label{vs1} \\ 
C_2^{\mbox{\scriptsize{(vac)}}} &=& \frac{e^4_q}{512 \pi^2} 
\frac{1}{m^4} \left[ 1 - \frac{4 \alpha_s}{3 \pi} \ln 
\frac{\Lambda}{m} \right] \ln \frac{\Lambda}{m} \, , 
\label{vs2} 
\end{eqnarray}
%%%%%%%%%%%%%%%%%%%%%%%%%%%%%%
and 
%%%%%%%%%%%%%%%%%%%%%%%%%%%%%%
\begin{eqnarray}
C_1^{\mbox{\scriptsize{(mat)}}} &=& \frac{3 e^2_q}{32 \pi^2} 
\frac{\mu_B^2}{9 m^2} \left[ 1 - \frac{ 4 \alpha_s}{3 \pi} \ln 
\frac{3 \Lambda}{\mu_B} \right] \, , 
\label{mu3} \\ 
C_2^{\mbox{\scriptsize{(mat)}}} &=& - \frac{e^4_q}{512 \pi^2} 
\frac{1}{m^4} \left[ 1 - \frac{4 \alpha_s}{3 \pi} \ln 
\frac{3 \Lambda^2}{m \mu_B} \right] \ln \frac{\mu_B}{3 m} \, . 
\label{mu4} 
\end{eqnarray}
%%%%%%%%%%%%%%%%%%%%%%%%%%%%%%
Note that $C_1^{\mbox{\scriptsize{(vac)}}} \, 
\rule[-2mm]{.14mm}{6.5mm} \raisebox{-1.85mm}{\scriptsize{$\; 
\alpha_s \to 0$}}$, $C_2^{\mbox{\scriptsize{(vac)}}} \, 
\rule[-2mm]{.14mm}{6.5mm} \raisebox{-1.85mm}{\scriptsize{$\; 
\alpha_s \to 0$}}$ and $C_1^{\mbox{\scriptsize{(mat)}}} \, 
\rule[-2mm]{.14mm}{6.5mm} \raisebox{-1.85mm}{\scriptsize{$\; 
\alpha_s \to 0$}}$ are \lq\lq exact'' (cf. Eqs. (\ref{mei}) and 
(\ref{cut})). To logarithmic accuracy, $\ln \mu_B / (3 m) \sim \ln 
\mu_B / m$. Nevertheless we choose $\ln \mu_B / (3 m)$ because the 
baryon number of the quark is $1/3$ (cf. Eq. (\ref{bunpu})). Adding 
both contributions, we obtain 
%%%%%%%%%%%%%%%%%%%%%%%%%%%%%%
\begin{eqnarray}
C_j &=& C_j^{{\mbox{\scriptsize{(vac)}}}} + 
C_j^{{\mbox{\scriptsize{(mat)}}}} = C_j^{(0)} + \frac{4 \alpha_s}{3 
\pi} C_j^{(1)} \;\;\;\;\;\;\;\;\;\;\;\; (j = 1, 2) \, , \\ 
C_1^{(0)} &=& - \frac{3 e_q^2}{32 \pi^2} \left( \ln 
\frac{\Lambda}{m} - \frac{\mu_B^2}{9 m^2} \right) \, , \\ 
C_1^{(1)} &=& - \frac{3 e_q^2}{32 \pi^2} \left[ \left( \ln 
\frac{\Lambda}{m} \right)^2 + \frac{\mu_B^2}{9 m^2} \ln \frac{3 
\Lambda}{\mu_B} \right] \, , \\ 
C_2^{(0)} &=& \frac{e_q^4}{512 \pi^2} \frac{1}{m^4} \ln \frac{3 
\Lambda}{\mu_B} \, , \\ 
C_2^{(1)} &=& - \frac{e_q^4}{512 \pi^2} \frac{1}{m^4} \left( \ln 
\frac{3 \Lambda}{\mu_B} \right)^2 \, . 
\label{end} 
\end{eqnarray}
%%%%%%%%%%%%%%%%%%%%%%%%%%%%%%

Several comments on the $O (g^2)$ contributions are in order. To 
logarithmic accuracy, the contributions $\langle \vec{\mu}' 
\rangle_2$ (Eq. (\ref{3-2})) and $\langle \vec{\mu}' \rangle_1'$ 
(Eq. (\ref{mei1})) are negligible. In other ward, the leading 
contributions displayed above come from $\langle \vec{\mu}' 
\rangle_1^{\mbox{\scriptsize{(self)}}}$, to which the responsible 
diagrams include the vacuum piece $\Sigma_{\mbox{\scriptsize{vac}}}$ 
of the quark self-energy part. We see the point in more detail. 
\begin{description} 
\item{On $C_1^{\mbox{\scriptsize{(vac)}}} \in \langle \vec{\mu}' 
\rangle_1'$:} As mentioned above, integration over $P$ (with $Q$ 
fixed) in Eq. (\ref{3-1}) converges. If the $P$-integration yields 
the term that is proportional to $\ln (\sqrt{- Q^2} /m)$, the 
remaining $Q$-integration yields $\ln^2 (\Lambda / m)$ term. This 
is, however, not the case. 
\item{On $C_2^{\mbox{\scriptsize{(vac)}}} \in \langle \vec{\mu}' 
\rangle_1'$:} $P$-integration (with $Q$ fixed) in Eq. (\ref{3-1}) 
yields the terms being proportional to $\ln (\sqrt{- Q^2} /m)$, 
which lead to the $\ln^2 (\Lambda / m)$ terms. Cancellation occurs 
among them and no $\ln^2 (\Lambda / m)$ term remains. 
\item{On $C_2^{\mbox{\scriptsize{(mat)}}} \in \langle \vec{\mu}' 
\rangle_1'$:} $Q$-integration using the regularized gluon propagator 
yields $\ln (\Lambda / m)$ terms. Cancellation occurs among those 
terms and then no $\ln (\Lambda / m)$ term remains. If the 
$Q$-integration yields the $\ln [(m^2 - P^2) / m^2]$ 
term, the remaining $P$-integration results in the leading $\ln^2 
(\mu_B / (3 m))$ contribution. Among the $Q$-integrated terms are 
the $\ln [(m^2 - P^2) / m^2]$ ones. Adding all of those, 
cancellation occurs again, so that no $\ln^2 (\mu_B / (3 m))$ term 
remains. 
\item{On $\langle \vec{\mu}' \rangle_2$:} Cancellation occurs 
between the leading contributions, so that there remains no leading 
contribution in $C_1^{\mbox{\scriptsize{(vac)}}}$, 
$C_1^{\mbox{\scriptsize{(mat)}}}$, 
$C_2^{\mbox{\scriptsize{(vac)}}}$ and 
$C_2^{\mbox{\scriptsize{(mat)}}}$. 
\end{description} 
%%%%%%%%%%%%%%%%%%%%%%%%%%%%%%%%%%%%%%%%%%%%%%%%%%%%%%%%
%%% SUB SUB %%%%%%%%%%%%%%%%%%%%%%%%%%%%%%%%%%%%%%%%%%%%%
%%%%%%%%%%%%%%%%%%%%%%%%%%%%%%%%
\subsection{Resummation} 
As has been mentioned above, the leading contributions (\ref{vs1}) 
- (\ref{mu4}) come from $\langle \vec{\mu}' 
\rangle_1^{\mbox{\scriptsize{(self)}}}$, for which the diagrams 
include $\Sigma_{\mbox{\scriptsize{vac}}}$. In the region $(4 
\alpha_s / 3 \pi) \ln \Lambda / m \sim 1$ and/or $(4 \alpha_s / 3 
\pi) \ln \mu_B / (3 m) \sim 1$, the expressions (\ref{vs1}) - 
(\ref{mu4}) are not reliable and an improvement is necessary through 
resummation of $\Sigma_{\mbox{\scriptsize{vac}}}$. To this end, it 
is convenient to start from Eqs. (\ref{mei}) and (\ref{meic}) and, 
on the RHS's of them, make the substitutions 
%%%%%%%%%%%%%%%%%%%%%%%%%%%%%%
\begin{eqnarray}
P & \to & P (1 - \Sigma_w) \equiv Z_2^{- 1} P (1 - \Sigma_w') \, , 
\nonumber \\ 
m & \to & m + \Sigma_m \equiv Z_2^{- 1} Z_m (m + \Sigma_m') \, . 
\label{cat} 
\end{eqnarray}
%%%%%%%%%%%%%%%%%%%%%%%%%%%%%%%%%%%%
Here $\Sigma_w = \mbox{Tr} 
(P\kern-0.1em\raise0.3ex\llap{/}\kern0.15em\relax 
\Sigma_{\mbox{\scriptsize{vac}}}) / (4 P^2)$, $\Sigma_m = \mbox{Tr} 
\Sigma_{\mbox{\scriptsize{vac}}} / 4$ and $Z_2$ $(Z_m)$ is the 
wave-function (mass) renormalization constant, 
%%%%%%%%%%%%%%%%%%%%%%%%%%%%%%
\begin{eqnarray*}
Z_2 &=& 1 - \frac{2 \alpha_s}{3 \pi} \ln \frac{\Lambda}{m} \, , \\ 
Z_m &=& 1 + \frac{2 \alpha_s}{\pi} \ln \frac{\Lambda}{m} \, . 
\end{eqnarray*}
%%%%%%%%%%%%%%%%%%%%%%%%%%%%%%%%%%%%
The substitutions (\ref{cat}) are not made for $m$ within $e_q / (4 
m)$ and for $p_0$ in $\theta (\mu_B / 3 - p_0)$ in Eq. (\ref{meic}). 
It should be noted that we are using the Feynman gauge here. 

The leading-logarithmic contributions (\ref{vs1}) - (\ref{mu4}) 
emerge from $Z_2$, $Z_m$ and the piece $\ln (m^2 - P^2) / m^2$ in 
$\Sigma_w'$ and $\Sigma_m'$. Then, to logarithmic accuracy, it is 
sufficient to resum the $\ln (m^2 - P^2) / m^2$ pieces in 
$\Sigma_w'$ and $\Sigma_m'$, which is gauge-parameter ($\eta$) 
independent. Thus, Eq. (\ref{cat}) may be simplified as 
%%%%%%%%%%%%%%%%%%%%%%%%%%%%%%
\begin{eqnarray}
P & \to & Z_2^{- 1} P \left( 1 - \frac{\alpha_s}{3 \pi} \ln 
\frac{m^2 - P^2}{m^2} \right) \, , \\ 
m & \to & Z_2^{- 1} Z_m m \left( 1 - \frac{4 \alpha_s}{3 \pi} 
\ln \frac{m^2 - P^2}{m^2} \right) \, . 
\label{ohosi} 
\end{eqnarray}
%%%%%%%%%%%%%%%%%%%%%%%%%%%%%%
Straightforward manipulation yields 
%%%%%%%%%%%%%%%%%%%%%%%%%%%%%%
\begin{eqnarray}
C_1^{\mbox{\scriptsize{(vac)}}} &=& - \frac{3 e^2_q}{32 \pi^2} 
Z_2^2 Z_m^2 \frac{1 - 5 \frac{2 \alpha_s}{3 \pi} \ln 
\frac{\Lambda}{m} + 7 \left( \frac{2 \alpha_s}{3 \pi} \ln 
\frac{\Lambda}{m} \right)^2}{\left( 1 - \frac{2 \alpha_s}{3 \pi} \ln 
\frac{\Lambda}{m}\right)^3} \ln \frac{\Lambda}{m} \, , 
\label{res1} \\ 
C_2^{\mbox{\scriptsize{(vac)}}} &=& - \frac{e^4_q}{1024 \pi} 
\frac{1}{m^4 \alpha_s} Z_2^4 \left[ 1 - \left( 1 - \frac{2 
\alpha_s}{3 \pi} \ln \frac{\Lambda}{m} \right)^{- 3} \right] \, , 
\nonumber \\ 
&& 
\label{res2} \\ 
C_1^{\mbox{\scriptsize{(mat)}}} &=& \frac{3 e^2_q}{32 \pi^2} 
\frac{\mu_B^2}{9 m^2} Z_2^2 \left[ 1 - \frac{ 2 \alpha_s}{3 \pi} 
\ln \frac{\mu_B}{3 m} \right]^{- 2} \, , 
\label{res3} \\ 
C_2^{\mbox{\scriptsize{(mat)}}} &=& \frac{e^4_q}{1024 \pi} 
\frac{1}{m^4 \alpha_s} Z_2^4 \left[ 1 - \left( 1 - \frac{2 
\alpha_s}{3 \pi} \ln \frac{\mu_B}{3 m} \right)^{- 3} \right] \, . 
\label{res4} 
\end{eqnarray}
%%%%%%%%%%%%%%%%%%%%%%%%%%%%%%
Adding both contributions, we obtain 
%%%%%%%%%%%%%%%%%%%%%%%%%%%%%%
\begin{eqnarray}
C_1 &=& C_1^{{\mbox{\scriptsize{(vac)}}}} + 
C_1^{{\mbox{\scriptsize{(mat)}}}} \nonumber \\ 
&=& - \frac{3 e_q^2}{32 \pi^2} Z_2^2 \left[ Z_m^2 \Omega^{- 1} 
\left( 7 - 9 \Omega^{- 1} + 3 \Omega^{- 2} \right) \ln 
\frac{\Lambda}{m} - \frac{\mu_B^2}{9 m^2} \Omega_m^{- 2} \right] 
\, , 
\label{c1} \\ 
C_2 &=& C_1^{{\mbox{\scriptsize{(vac)}}}} + 
C_2^{{\mbox{\scriptsize{(mat)}}}} \nonumber \\ 
&=& \frac{e_q^4}{1024 \pi} \frac{1}{m^4 \alpha_s} Z_2^4 \left( 
\Omega^{- 3} - \Omega^{- 3}_m \right) \, , 
\label{endfin} 
\end{eqnarray}
%%%%%%%%%%%%%%%%%%%%%%%%%%%%%%
where 
%%%%%%%%%%%%%%%%%%%%%%%%%%%%
\begin{equation} 
\Omega = 1 - \frac{2 \alpha_s}{3 \pi} \ln \frac{\Lambda}{m} \left( = 
Z_2 \right)\, , \;\;\;\;\;\;\;\;\;\;
\Omega_m = 1 - \frac{2 \alpha_s}{3 \pi} \ln \frac{\mu_B}{3 m} \, . 
\label{omega} 
\end{equation} 
%%%%%%%%%%%%%%%%%%%%%%%%%%%%%%%%%%
%%%%%%%%%%%%%%%%%%%%%%%%%%%%%%
%%%%%%%%%%%%%%%%%%%%%%%%%%%%%
\section{Effective potential} 
In this section, we study the structure of the effective potential 
$V ( \langle \vec{\mu} \rangle)$. The \lq\lq generating function'' 
$W (\vec{B}^{\, 2})$ is related to $\langle \vec{\mu}' \rangle$, 
Eq. (\ref{mu}), through 
%%%%%%%%%%%%%%%%%%%%%%%%%%%%%%%%%%%%%%%%%%%%%%
\begin{equation} 
W \left( \vec{B}^{\, 2} \right) = \int_0^{\vec{B}} \langle 
\vec{\mu}' \, \rangle d \vec{B} = \frac{C_1}{2} \vec{B}^{\, 2} + 
\frac{C_2}{4} \left( \vec{B}^{\, 2} \right)^2 
\label{kikuchan}
\end{equation} 
%%%%%%%%%%%%%%%%%%%%%%%%%%%%%%%%%%%%%%%%%%%%%%%%%%%%%%%%%
with $C_1$ and $C_2$ as in Eqs. (\ref{c1}) and (\ref{endfin}), 
respectively. $V \left( \langle \vec{\mu}' \rangle \right)$ is 
defined by the following Legendre transformation on $W \left( 
\vec{B}^{\, 2} \right)$: 
%%%%%%%%%%%%%%%%%%%%%%%%%%%%%%%%%%%%%%%%%%%%%%
\begin{eqnarray} 
V \left( \langle \vec{\mu}' \rangle \right) & = & \langle \vec{\mu}' 
\rangle \vec{B} - W \left( \vec{B}^{\, 2} \right) \, , 
\label{yuukou} \\ 
\frac{d V \left( \langle \vec{\mu}' \rangle \right)}{d \langle 
\vec{\mu}' \rangle} &=& \vec{B} \, , 
\label{4.2d} 
\end{eqnarray} 
%%%%%%%%%%%%%%%%%%%%%%%%%%%%%%%%%%%%%%%%%%%%%%%%%%%%%%%%%
where $\vec{B}$ $\left( \equiv \vec{B} \left( \langle \vec{\mu}' 
\rangle \right) \right)$ is related to $\langle \vec{\mu}' \rangle$ 
through Eq. (\ref{mu}). 

When one is interested in $V \left( \langle \vec{\mu}' \rangle 
\right)$ with small $\langle \vec{\mu}' \rangle$, it is sufficient 
to obtain $\vec{B} \left( \langle \vec{\mu}' \rangle \right)$ in 
powers of $\langle \vec{\mu}' \rangle$. For obtaining more accurate 
information, however, we solve Eq. (\ref{mu}) exactly. 

As will be seen below, in the region of our interest, $C_2 > 0$. For 
convenience, let us introduce the dimensionless quantities, 
%%%%%%%%%%%%%%%%%%%%%%%%%%%%%%%%%%%%%%%%%%%%
\begin{eqnarray} 
\vec{M} & \equiv & \frac{3}{2} \sqrt{\frac{3 C_2}{| C_1 |^3}} 
\langle \vec{\mu}' \rangle = M \frac{\vec{M}}{|M|} \,\,\,\,\, (M 
\geq 0) \, , 
\label{emu} \\ 
\vec{\cal B} & \equiv & \sqrt{\frac{3 C_2}{|C_1|}} \vec{B} 
= {\cal B} \frac{\vec{M}}{|M|} \, , \\ 
{\cal V} (M) & \equiv & \frac{C_2}{| C_1 |^2} \, V \left( \langle 
\vec{\mu}' \rangle \right) \, . 
\end{eqnarray} 
%%%%%%%%%%%%%%%%%%%%%%%%%%%%%%%%%%%%%%%%%%%%
Eqs. (\ref{mu}), (\ref{yuukou}) and (\ref{4.2d}) turn out, in 
respective order, to 
%%%%%%%%%%%%%%%%%%%%%%%%%%%%%%%%%%%%%%%%%%%%
\begin{eqnarray} 
&& {\cal B}^3 (M) + 3 \epsilon (C_1) {\cal B} (M) - 2 M = 0 \, , 
\label{kiso} \\ 
&& {\cal V} (M) = \frac{2}{9} M \, {\cal B} (M) - \frac{1}{6} 
\epsilon (C_1) {\cal B}^2 (M) - \frac{1}{36} {\cal B}^4 (M) \, , 
\label{redeq} \\ 
&& \frac{d {\cal V} (M)}{d M} = \frac{2}{9} {\cal B} (M) \, , 
\label{atsumi}
\end{eqnarray} 
%%%%%%%%%%%%%%%%%%%%%%%%%%%%%%%%%%%%%%%%%%%%
where $\epsilon (C_1) = C_1 / | C_1 |$. 

Solving Eq. (\ref{kiso}), we obtain the following expressions for 
${\cal B} (M)$.  
\subsubsection*{$C_1 > 0$}
%%%%%%%%%%%%%%%%%%%%%%%%%%%%%
\begin{equation}
{\cal B} (M) = \left( \sqrt{1 + M^2} + M \right)^{1 / 3} - \left( 
\sqrt{1 + M^2} - M \right)^{1 / 3} \, . 
\label{soll}  
\end{equation}
%%%%%%%%%%%%%%%%%%%%%%%%%%%%%%
From Eqs. (\ref{atsumi}) and (\ref{soll}), we see that 
%%%%%%%%%%%%%%%%%%%%%%%%%%%%%%
\begin{eqnarray*}
\frac{d {\cal V} (M)}{d M} & > & 0 \;\;\;\; \mbox{for} \; M > 0 \, , 
\\ 
\frac{d {\cal V} (M)}{d M} & = & 0 \;\;\;\; 
\mbox{at} \; M = 0 \, . 
\end{eqnarray*}
%%%%%%%%%%%%%%%%%%%%%%%%%%%%%%%%%
\subsubsection*{$C_1 < 0$} 
(1) $M \geq 1$ 

\hspace*{3ex} Solution (s1): 
%%%%%%%%%%%%%%%%%%%%%%%%%%%%%
\begin{eqnarray*}
{\cal B} (M) & = & \left( M + \sqrt{M^2 - 1} \right)^{1 / 3} + \left( 
M - \sqrt{M^2 - 1} \right)^{1 / 3} \, , \\ 
\frac{d {\cal V} (M)}{d M} & > & 0 \, . 
\end{eqnarray*}
%%%%%%%%%%%%%%%%%%%%%%%%%%%%%%%%%

(2) $M < 1$ 

\hspace*{3ex} Solution (s2): 
%%%%%%%%%%%%%%%%%%%%%%%%%%%%%%%%%%%%%%%%%%%%%%%%%%%%%%%%%%%%%
\begin{eqnarray*} 
{\cal B} (M) & = & 2 \cos \left( \frac{\arccos M}{3} \right) \, , \\ 
\frac{d {\cal V} (M)}{d M} &>& 0 \, . 
\end{eqnarray*} 
%%%%%%%%%%%%%%%%%%%%%%%%%%%%%%%%%%%%%%%%%%%%%%%%%%%%%%%%%%%%%%%

\hspace*{3ex} Solution (s3): 
%%%%%%%%%%%%%%%%%%%%%%%%%%%%%%%%%%%%%%%%%%%%%%%%%%%%%%%%%%%%%
\begin{eqnarray*} 
{\cal B} (M) & = & 2 \cos \left( \frac{4 \pi + \arccos M}{3} \right) 
\, , 
\label{ai} 
\\ 
\frac{d {\cal V} (M)}{d M} &<& 0 \;\;\;\; \mbox{for} \; 0 < M < 1 
\, , \\ 
\frac{d {\cal V} (M)}{d M} &=& 0 \;\;\;\; \mbox{at} \; M = 0 \, . 
\end{eqnarray*} 
%%%%%%%%%%%%%%%%%%%%%%%%%%%%%%%%%%%%%%%%%%%%%%%%%%%%%%%%%%%%%%%

\hspace*{3ex} Solution (s4): 
%%%%%%%%%%%%%%%%%%%%%%%%%%%%%%%%%%%%%%%%%%%%%%%%%%%%%%%%%%%%%
\begin{eqnarray*} 
{\cal B} (M) & = & 2 \cos \left( \frac{2 \pi + \arccos M}{3} \right) 
\, , \\ 
\frac{d {\cal V} (M)}{d M} &<& 0 \, . 
\end{eqnarray*} 
%%%%%%%%%%%%%%%%%%%%%%%%%%%%%%%%%%%%%%%%%%%%%%%%%%%%%%%%%%%%%%%

In Fig. 2, we display ${\cal V} (M)$ in the case of $C_1 > 0$. The 
curve \lq\lq 0'' is ${\cal V} (M)$ with Eq. (\ref{soll}) for 
${\cal B} (M)$. We see that, when $C_1 > 0$, the quark matter is in 
the normal phase. 

In Fig. 3(a), we depict ${\cal V} (M)$ for $C_1 < 0$. The curves 
(s1), (s2), (s3) and (s4) show ${\cal V} (M)$ corresponding, in 
respective order, to the solutions (s1), (s2), (s3) and (s4) for 
${\cal B} (M)$. The actual effective potential ${\cal V} (M)$ is as 
in Fig. 3(b). From this figure, we see that, when $C_1 < 0$, ${\cal 
V} (M)$ is minimum at $M = 1$ or $\langle \vec{\mu}' \rangle = 
\sqrt{4 |C_1|^3 / (27 C_2)}$, which means that the spontaneous 
magnetization occurs. 

Here we discuss the effect of the higher-order (in $\vec{B}$) 
contribution to $\langle \vec{\mu}' \rangle$ in Eq. (\ref{mu}), 
%%%%%%%%%%%%%%%%%%%
\begin{equation}
\langle \vec{\mu}' \rangle = \left[ C_1 + C_2 \vec{B}^{\, 2} + C_3 
( \vec{B}^{\, 2} )^2 \right] \vec{B} \, . 
\label{alp}
\end{equation}
%%%%%%%%%%%%%%%%%%%
Eqs. (\ref{kiso}) and (\ref{redeq}) turn out to be 
%%%%%%%%%%%%%%%%%%%%%%%%%%%%%%%%%%%%%%%%%%%%
\begin{eqnarray} 
&& {\cal B}^3 (M) + 3 \epsilon (C_1) {\cal B} (M) + 
\frac{|C_1| C_3}{3 C_2^2} {\cal B}^5 (M) - 2 M = 0 \, , 
\label{kisoa} \\ 
&& {\cal V} (M) = \frac{2}{9} M \, {\cal B} (M) - \frac{1}{6} 
\epsilon (C_1) {\cal B}^2 (M) - \frac{1}{36} {\cal B}^4 (M) 
- \frac{|C_1| C_3}{162 C_2^2} {\cal B}^6 (M) \, . 
\label{redeqa} 
\end{eqnarray} 
%%%%%%%%%%%%%%%%%%%%%%%%%%%%%%%%%%%%%%%%%%%%
From these equations, we see first of all that, near the phase 
transition point, $C_1 \simeq 0$, the effects of the \lq\lq $C_3$ 
term'' is very small. We will compute ${\cal V} (M)$ for the 
following two choices for $|C_1| C_3 / C_2^2$: 
\[
\frac{|C_1| C_3}{C_2^2} = -2, \;\, + 2 \, .  
\]
The curve \lq\lq +'' (\lq\lq -'') in Fig. 2 depicts ${\cal V} (M)$ 
with $|C_1| C_3 / C_2^2 = + 2$ $(- 2)$. We see that the effect of 
the \lq\lq $C_3$ term'' in Eq. (\ref{alp}) on ${\cal V} (M)$ is 
small and the system is in the normal phase. Fig. 3(c) depicts 
${\cal V} (M)$ in the case of $C_1 < 0$. For $|C_1| C_3 / C_2^2 = + 
2$, the behavior of ${\cal V} (M)$ is qualitatively the same as that 
for $C_3 = 0$ and the spontaneous magnetization takes place with 
$\langle \vec{\mu}' \rangle \simeq 0.84 \sqrt{4 |C_1|^3 / (27 
C_2)}$. On the other hand, for $|C_1| C_3 / C_2^2 = - 2$, ${\cal V} 
(M)$ decreases monotonically as $M$ increases. If we take this 
result at face value, the system is in the ferromagnetic phase. 

Recalling the fact that $C_1$ is a function of $\Lambda$ and 
$\mu_B$, we summarize the above observations: 
\begin{itemize}
\item In the region of $\mu_B$ where $C_1 (\Lambda, \mu_B) > 0$, the 
system is in the normal phase. 
\item In the region where $C_1 (\Lambda, \mu_B) < 0$, the system 
is in the ferromagnetic phase. 
\item At $\mu_B = \mu_{Bc}$, where $C_1 (\Lambda, \mu_{Bc}) = 0$, 
a magnetic phase transition between the above two phases tales 
place. 
\end{itemize}
%%%%%%%%%%%%%%%%%%%%%%%%%%%%%%%%%%
%%%%%%% SEC %%%%%%%%%%%%%%%%%%%%%%
%%%%%%%%%%%%%%%%%%%%%%%%%%%%%
\section{Numerical Analysis}
In this section, on the basis of the observation in Sec. 4, we study 
the magnetic property of the quark matter through the numerical 
analysis in the region $0.6$ GeV $\leq \mu_B \leq 4$ GeV. Let start 
with two observations.
\begin{itemize} 
\item As the renormalization scale $\mu_R$ (for the running coupling 
constant $\alpha_s (\mu_R)$ and the running mass $m (\mu_R)$), we 
choose $\mu_R = \mu_B$. There is an ambiguity on the choice of 
$\mu_R$. For computing the pressure of the quark-gluon plasma (QGP) 
at temperature $T$ and $\mu_B = 0$, $\mu_R \simeq 2 \pi T$ is chosen 
in the literature (see, e.g., Ref. 9)). The so-called thermal 
mass of a zero mass quark in QGP with finite $T$ and $\mu_B$ is 
$m_{\mbox{\scriptsize{th}}} \propto T^2 + \mu_B^2 / (3 \pi^2)$. 
These observations suggest the choice, $\mu_R = 2 \pi [T^2 + \mu_B^2 
/ (3 \pi^2)]^{1/ 2}$ $\stackrel{T \to 0}{\longrightarrow}$ $2 \pi 
\times (\mu_B / (\sqrt{3} \pi)) \simeq 1.15 \mu_B$, which is not 
very different from $\mu_B$. 
\item Following Ref. 10), we identify the effective potential $V$ 
computed above with cutoff $\Lambda$ with {\em the one renormalized 
at the renormalization scale $\mu_R = \Lambda$ in QCD}. 
\end{itemize} 
On the basis of these observations, as a crude approximation, we 
choose $\Lambda = \mu_B$. Then, from Eqs. (\ref{endfin}) and 
(\ref{omega}), $C_2 (\mu_B) > 0$, provided that $\Omega (\mu_B)> 0$, 
which is the case in the region of our interest. 

For the QCD $\alpha_s$ ($= g^2 / (4 \pi)$), we take the running 
one $\alpha_s (\mu_B)$ in the $\overline{\mbox{MS}}$ scheme with 
$\alpha_s (\mu_B = 1.2 \; \mbox{GeV}) = 0.39$. In Fig. 4, $\alpha_s 
(\mu_B)$ is depicted in the relatively small $\mu_B$ region, $600$ 
MeV $\leq \mu_B \leq 700$ MeV. For the quark masses, we take 1-loop 
running masses 
%%%%%%%%%%%%%%%%%%%
\[
m (\mu_R) = m (\mu_R') \left( \frac{\alpha_s (\mu_R)}{\alpha_s 
(\mu_R')} \right)^{2 / (\pi \beta_0)} \;\;\;\;\;\;\;\; (\beta_0 = 
(11 - 2 n_f / 3) / (2 \pi) ) \, , 
\]
%%%%%%%%%%%%%%%%%%%%%
with 
%%%%%%%%%%%%%%%%%%%%%
\begin{eqnarray*}
m_u (\mu_R = 2 \; \mbox{GeV}) & = & 3 \; \mbox{MeV/c}^2, 
\;\;\;\;\;\; m_d (\mu_R = 2 \; \mbox{GeV}) = 6.8 \; 
\mbox{MeV/c}^2, \\ 
m_s (\mu_R = 2 \; \mbox{GeV}) & = & 118 \; \mbox{MeV/c}^2, 
\;\;\;\;\;\; m_c (\mu_R = 1.2 \; \mbox{GeV}) = 1.2 \; 
\mbox{GeV/c}^2 \, . 
\end{eqnarray*}
%%%%%%%%%%%%%%%%%%%%%%%%%%%%%%%
$n_f$ is the number of quark flavors and the quark with mass $m_q$ 
is counted for $m_q < \mu_R$ ($ = \mu_B$). For the QED fine 
structure constant $\alpha$, we take the 1-loop running one. For 
computing it, we have used the above-mentioned quark masses and, for 
lepton masses, we have used $m_e = 0.5$ MeV/c$^2$, $m_\mu = 106$ 
MeV/c$^2$ and $m_\tau = 1.78$ GeV/c$^2$. 

The relation between the baryon-number density $\rho$ and the 
baryonic chemical potential $\mu_B$ is 
%%%%%%%%%%%%%%%%%%%%%%%%%%%%%%%%%%%
\begin{equation}
\rho (\mu_B) = 3 \times 2 \times \frac{1}{6 \pi^2} \sum_{i = u, d, 
s, c} \theta \left( \frac{\mu_B}{3} - m_i (\mu_B) \right) \left( 
\frac{\mu_B^2}{9} - m_i^2 (\mu_B) \right)^{3 / 2} \, , 
\label{atsumi1} 
\end{equation}
%%%%%%%%%%%%%%%%%%%%%%%%%
where \lq\lq 3'' is the number of colors and \lq\lq 2'' comes from 
the spin degree of freedom. We do not take into account the QCD 
correction to the relation (\ref{atsumi1}) except that the running 
masses have been used. The nuclear density at the center of a nuclei 
is $\rho_0 \simeq 0.17$ fm$^{- 3}$. The density of neutron stars are 
in the range, $\rho = 10^{- 3} \rho_0 \sim 10 \rho_0$. For a guide 
of eyes, the relation between $\rho / \rho_0$ and $\mu_B$ is shown 
in Figs. 5(a) and (b), and the ratio $\rho_i / \rho$ $(i = u, d, s, 
c)$ is shown in Fig. 5(c). 
%%%%%%%%%%%%%%%%%%%%%%%%%%%%%%%%%%
%%%%%%%%%%%%%%% SUBSUB %%%%%%%%%%%%%%%%%%%
\subsubsection*{Quarks} 
As mentioned above, $C_2 (\mu_B) > 0$ in the region of interest. 
Numerical analysis shows that 
%%%%%%%%%%%%%%%%%%%%%%%%%%%%%
\[
\left\{ \begin{array}{ll}
C_1 (\mu_B) < 0  & \; \mbox{for} \; \mu_B < \mu_{B c} \, , \\ 
C_1 (\mu_B) = 0  & \; \mbox{for} \; \mu_B = \mu_{B c} \, , \\ 
C_1 (\mu_B) > 0  & \; \mbox{for} \; \mu_B > \mu_{B c} \, , 
\end{array}{}
\right. 
\]
%%%%%%%%%%%%%%%%%%%%%%%%%%%%%
with $\mu_{B c} \simeq 641$, $625$ and $693$ MeV for, in respective 
order, $u$, $d$ and $s$ quarks. In Fig. 6, $C_1 (\mu_B)$ is depicted 
against $\mu_B$ for $600$ MeV $\leq \mu_B \leq 700$ MeV, where the 
coefficient $d_1 = 1, 50$ and $2 \times 10^{4}$ for $u$, $d$ and $s$ 
quarks, respectively. Vertical dashed lines indicate the phase 
transition points. We see that, at high-density region, $\mu_B > 
\mu_{Bc}$, $C_1 > 0$ and the quarks are in the normal phase. As the 
density is lowered, the quarks undergo a magnetic phase transition 
($C_1 = 0$), at the critical density, into a ferromagnetic phase 
($C_1 < 0$). We see from Fig. 3(b) (also Fig. 3(c)) that, in the 
ferromagnetic phase, there arises spontaneous magnetization whose 
magnitude is $| \langle \vec{\mu}' \rangle | =  \sqrt{4 |C_1|^3 / 
(27 C_2)} \, M \simeq \sqrt{4 |C_1 |^3 /(27 C_2)}$ $(\equiv 
\langle \mu' \rangle)$ (cf. Eq. (\ref{emu})). We depict $\langle 
\mu' \rangle$ in Fig. 7, where $d_2 = 10^{- 4}, 2 \times 10^{- 3}$ 
and $0.1$ for $u$, $d$, and $s$ quarks respectively. 

$C_1$ is the quantity that determines the phase of the system. Eq. 
(\ref{res1}) with $\Lambda = \mu_B$ tells us that the vacuum-sector 
contribution to $C_1$ is negative, $C_1^{\mbox{\scriptsize{(vac)}}} 
< 0$, in the region of our interest. While, from Eq. (\ref{res3}), 
$C_1^{\mbox{\scriptsize{(mat)}}} (\mu_B)$ is positive and an 
increasing function of $\mu_B$. The effect of the QCD interaction 
does not change the signs of $C_1^{\mbox{\scriptsize{(vac)}}}$ and 
$C_1^{\mbox{\scriptsize{(mat)}}}$. At low densities 
$C_1^{\mbox{\scriptsize{(mat)}}}$ is so small that $C_1 = 
C_1^{\mbox{\scriptsize{(vac)}}} + C_1^{\mbox{\scriptsize{(mat)}}} < 
0$. As $\mu_B$ is increased $C_1^{\mbox{\scriptsize{(mat)}}}$ 
increases and we arrive at $C_1 = 0$ at $\mu_B = \mu_{Bc}$. As 
$\mu_B$ further increases, $C_1$ turns out to be positive. To 
understand the behavior of $C_1^{\mbox{\scriptsize{(mat)}}} 
(\mu_B)$, let us recall that the hamiltonian density describing the 
interaction of the system with the external magnetic field is ${\cal 
H}_{\mbox{\scriptsize{int}}} = - \vec{\mu}' \cdot \vec{B}$. This 
means that the configuration $\vec{\mu}' \parallel \vec{B}$ is 
energetically favorable, which reflects on 
$C_1^{\mbox{\scriptsize{(mat)}}} \, \rule[-2mm]{.14mm}{6.5mm} 
\raisebox{-1.85mm}{\scriptsize{$\; \alpha_s = 0$}} > 0$, Eq. 
(\ref{mu3}) or (\ref{res3}) (cf. Eq. (\ref{mu})). 

Several observations are in order here. 
\begin{itemize}
\item In spite of the fact that $m_u < m_d < m_s$, we have obtained 
the curious result, $(\mu_{Bc})_d < (\mu_{Bc})_u < (\mu_{Bc})_s $. 
This means that, as $\mu_B$ is lowered from the normal-phase region, 
the heaviest s quarks are magnetized first, then follows the 
lightest u quarks and finally d quarks start are magnetized. 
\item The magnitude of the magnetization is larger for the lighter 
quark. This is a reflection of the fact that $\vec{\mu}' \propto 1 / 
m$. 
\item The phase transition points $\mu_{Bc} \simeq 641, 625$ and 
$693$ MeV for, in respective order, $u$, $d$ and $s$ quarks 
correspond 
to the densities $\rho \simeq 1.62 \rho_0$, $1.47 \rho_0$ and $2.20 
\rho_0$ with $\rho_0$ the nuclear density. These values for $\rho$ 
are near to the nuclear density $\rho_0$ and then our results may be 
relevant to the quark star. 
\item As seen from Fig. 4, $\alpha_s (\mu_B)$ is not very small at 
$600$ MeV $\leq \mu_B \leq 700$ MeV. Then, before arriving 
at definite conclusion, higher-order (in $\alpha_s$) contributions 
should be taken into account.\footnote{In this relation, we recall 
the fact that, for a dense electron system, the higher-order 
contributions substantially changes \cite{iida} the magnetic phase 
diagram when compared to the calculations using the random-phase 
approximation.} 
\end{itemize}
Here we mention the case with the order parameter $\vec{\mu} = 2 
\vec{\mu}'$ (see Appendix A). It can readily be seen that $C_1 = 4 
C_1 (\mbox{Eq. (\ref{c1})}) / 4$ and $C_2 = 16 C_2 (\mbox{Eq. 
(\ref{endfin})})$. The phase transition point $\mu_{B c}$ does 
not change. 
%%%%%%%%%%%%%%%%%%%%%%%%%%%%%%%%%%%%%%%%%%%%%
\subsubsection*{Leptons} 
Our primary interest is to study the magnetic property of the 
quark matter, like quark star. Such a system is electrically 
neutral, so that there exists leptons. From the charge neutrality, 
we obtain the relation between the baryonic chemical potential 
$\mu_B$ and the leptonic chemical potential $\mu_L$ being conjugate 
to the lepton number. We see that, for $0.6 \; \mbox{GeV} \leq \mu_B 
\leq 4 \; \mbox{GeV}$, $\mu_L$ covers the range $174 \; \mbox{MeV} 
\leq \mu_L \leq 969 \; \mbox{MeV}$. In Fig. 8(a), we depict the 
relation between them in the region $0.6 \; \mbox{GeV} \leq \mu_B 
\leq 3 \; \mbox{GeV}$, and, in Fig. 8(b), the ratio $\rho_i / \rho$ 
$(i = e, \mu)$ is shown in the region $174 \; \mbox{MeV} \leq 
\mu_L \leq 300 \; \mbox{MeV}$. An abrupt increase in $\mu_L$ at 
$\mu_B \simeq 2.9$ GeV is due to the opening of the c quark channel, 
$m_c (\mu_B \simeq 2.9 \; \mbox{GeV}) \simeq 2.9 / 3$ GeV/c$^2$. 

For the QED coupling constant, we take $\alpha = 1 / 136$ and, for 
the lepton masses, we take the values cited at the beginning of Sec. 
5. Numerical analysis shows that, in the region $174 \; \mbox{MeV} 
\leq \mu_L \leq 969 \; \mbox{MeV}$, the leptons are in the normal 
phase. 
%%%%%%%%%%%%%%%%%%%%%%%%%%%%%%%%%%
%%%% SEC %%%%%%%%%%%%%%%%%%%%%%%%%%%%%%
%%%%%%%%%%%%%%%%%%%%%%%%%%%%%%%%%%
\section{Summary and Outlook}
%%%%%%%%%%%%%%%%%%%%%%%%%%%%%%%
In this paper, we have studied the magnetic property of the quark 
matter through evaluating the effective potential $V (\langle 
\vec{\mu}' \rangle)$ for the magnetic moment $\vec{\mu}'$ of a 
quark. We have found that, at low densities $\rho < \rho_c$, the 
quarks are in the ferromagnetic phase. For the $u$, $d$ and $s$ 
quarks, $\rho_c \simeq 1.62 \rho_0$, $1.47 \rho_0$ and $2.20 \rho_0$ 
($\rho_0$ the nuclear density), respectively. The phase transition 
occurs at $\rho = \rho_c$ and, at high densities $\rho > \rho_c$, 
the quarks turn out to be in the normal phase. 

We have also studied the case of leptons in the region $174 \; 
\mbox{MeV} \leq \mu_L \leq 969 \; \mbox{MeV}$ ($\mu_L$ the 
leptonic chemical potential), which corresponds to $0.6 \; 
\mbox{GeV} \leq \mu_B \leq 4 \; \mbox{GeV}$. We have found 
that both $e$ and $\mu$ are in the normal phase. 

We list the several points which are left for future study. 
\begin{enumerate} 
\item 
When more than two kinds of particles are in ferromagnetic phases, 
for determining the relative direction(s) of their magnetizations, 
interactions between them 
should be taken into account. 
\item Study of the quark matter at finite temperature and density. 
\item Analysis in the region \lq CSC' in Fig. 1, where several 
phases coexist. 
\item Undoing the assumption made after Eq. (\ref{or}) in Appendix 
A. 
\item Improvements of the approximation. 
\begin{description} 
\item{a)} Undoing the leading logarithmic approximation (\ref{app}). 
\item{b)} Resummations for the (quark and gluon) propagators and the 
quark-gluon vertex. 
\item{c)} Computation of the higher-order contributions. This is 
particularly important for studying the ferromagnetic phase, since 
$\alpha_s (\mu_B)$ is not very small there. 
\item{d)} Resummation of, e.g., the ladder diagrams. From thus 
obtained effective potential and the renormalization-group formula, 
one can expect to obtain (approximately) $\Lambda$-independent 
result. (For the case of chiral and diquark condensations, see, 
e.g., Ref. 10).) 
\end{description} 
\item Computation of the effective action, which allows us to study 
the dynamical aspect of the system. 
\item Study of nonequilibrium quark matter, which allows one to 
deal with the space-time evolution of the system under a given 
initial data. 
\end{enumerate} 
%%%%%%%%%%%%%%%%%%%%%%%%%%%%%%%
\section*{Acknowledgments}
The author thanks the useful discussion at the Workshop on Thermal 
Field Theories and their Applications, held at the Yukawa Institute 
for Theoretical Physics, Kyoto, Japan, 9 - 11 August, 2004. 
%%%%%%%%%%%%%%%%%%%%%%%%%%%%%%%%%%%%%%%%%%%%%%%%%%%%%%%%%%%%%%
%%%%%%%%%%%%%%%%%%%%%%%%%%%%%%%
\begin{appendix} %%%%%%%%%%%%%%%%%%%%%%%%%%%%%%%
\setcounter{equation}{0}
\setcounter{section}{1}
\section*{Appendix A Order parameter} 

We start with the action that describes the interaction 
between a quark (with charge $e_q$) and uniform external 
magnetic field $\vec{B}$: 
%%%%%%%%%%%%%%%%%%%%%%%%%%%%%%%
\begin{eqnarray*}
\int d^{\, 4} x \, {\cal L}_{\mbox{\scriptsize{ext}}} & = & e_q 
\int d^{\, 4} x \, \bar{\psi} \vec{\gamma} \cdot \vec{A} \psi 
\nonumber \\ 
& = & \frac{e_q}{2} \int d^{\, 4} x \, \bar{\psi} \vec{\gamma} \cdot 
(\vec{B} \times \vec{r}) \psi \, , 
\label{appe} 
\end{eqnarray*}
%%%%%%%%%%%%%%%%%%%%%%%%%%%%%%%%%%
where $\psi$ and $\bar{\psi}$ are the quark fields in the 
interaction representation. Here the color index is suppressed. 
Using the Dirac equation, we have 
%%%%%%%%%%%%%%%%%%%%%%%%%%%%%%%%%%%
%\begin{eqnarray*}
\[
0 = \bar{\psi} (x) \left(- i \stackrel{\leftarrow}{
\partial\mbox{\hspace*{0.3mm}}\kern-0.1em\raise0.3ex\llap{/}
\kern0.15em\relax} - m \right) \vec{\gamma} \cdot \vec{A} (x) 
\psi(x) + \bar{\psi} (x) \vec{\gamma} \cdot \vec{A} (x) \left( i 
\partial\mbox{\hspace*{0.3mm}}\kern-0.1em\raise0.3ex\llap{/}
\kern0.15em\relax - m \right) \psi(x) \, , 
\]
%\end{eqnarray*}
%%%%%%%%%%%%%%%%%%%%%%%%%%%%%%%%%%%
which yields 
%%%%%%%%%%%%%%%%%%%%%%%%%%%%%%%%%%%
\[ 
2 m \bar{\psi} \vec{\gamma} \cdot \vec{A} \psi = \bar{\psi} 
\left[ - i \vec{A} \cdot \vec{\nabla} 
+ A^i \partial_\mu \sigma^{i \mu} + i 
\stackrel{\leftarrow}{\partial_i} A^i - 
\stackrel{\leftarrow}{\partial_\mu} \sigma^{\mu i} A^i \right] \psi 
\, . 
\] 
%%%%%%%%%%%%%%%%%%%%%%%%%%%%%%%%%%%
Using $\vec{A} = \vec{B} \times \vec{r} / 2$, we finally obtain 
%%%%%%%%%%%%%%%%%%%%%%%%%%%%%%%%%%%
\begin{eqnarray} 
\int d^{\, 4} x \, {\cal L}_{\mbox{\scriptsize{ext}}} &=& 
\frac{e_q}{4 m} \int d^{\, 4} x \, 
\bar{\psi} \left[ - 2 i \vec{B} \cdot (\vec{r} \times \vec{\nabla}) 
+ \sigma^{k i} \epsilon^{i j k} B^j \right] \psi \nonumber \\ 
&=& \frac{e_q}{2 m} \int d^{\, 4} x \left[ - i \vec{r} \times 
\vec{\nabla} + \vec{\sigma} \right] \cdot \vec{B} \psi \, . 
\label{or} 
\end{eqnarray} 
%%%%%%%%%%%%%%%%%%%%%%%%%%%%%%%%%%%%%%%%

We take a statistical ensemble of the systems, each of which has 
vanishing total angular momentum. We further assume that, in each 
system, the total angular momentum of quarks and that of antiquarks 
vanish separately. Incidentally, the antiquarks exist as the virtual 
particles, through repetition of the $q \bar{q}$-pair productions 
(from the vacuum) and the pair annihilations. Then, Eq. (\ref{or}) 
turns out to 
%%%%%%%%%%%%%%%%%%%
\begin{eqnarray*}
\int d^{\, 4} x \, {\cal L}_{\mbox{\scriptsize{ext}}} & = & \int 
d^{\, 4} x \, \vec{\mu}' \cdot \vec{B} \, , \nonumber \\ 
\vec{\mu}' &=& \frac{\vec{\mu}}{2} = \frac{1}{2} \left( \frac{e_q}{2 
m} \right) \bar{\psi} \vec{\sigma} \psi = 
\frac{e_q}{4 m} \bar{\psi} \gamma_5 \gamma_0 
\vec{\gamma} \psi \, . 
\end{eqnarray*}
%%%%%%%%%%%%%%%%%%%%%
As the order parameter, we adopt $\vec{\mu}'$. 

It is to be noted that if we ignore the contribution from the 
orbital angular momentum in Eq. (\ref{or}), we obtain 
$\vec{\mu}$ $(= 2 \vec{\mu}')$ for the order parameter. 
%%%%%%%%%%%%%%%%%%%%%%%%%%%%%%%%%%%%%%
%%%%%%%%%%%%%%%%%%%%%%%%%%%%%%%%%%%
\end{appendix} %%%%%%%%%%%%%%%%%%%%%%%%%%%%%%%
%%%%%%%%%%%%%%%%%%%%%%%%%%%%%%%%%%
%%%%%% REFERENCES %%%%%%%%%%%%%%%%
%%%%%%%%%%%%%%%%%%%%%%%%%%%%%%%%%%

%%%%%%%%%%%%%%%%%%%%%%%%%%%%%%%%%%%%%%%%%%%%%%%%%%%%%%%%%%%%%
%%%%%% FIG %%%%%%%%%%%%%%%%%%%%%%%%%%%%%%%%%%%%%
%%%%%%%%%%%%%%%%%%%%%%%%%%%%%%%%%%%%%
\newpage 
\begin{description} 
\item{Fig. 1.} Sketch of the QCD phase diagram. \lq $T$' is the 
temperature and \lq $\mu_B$' is the baryonic chemical potential. \lq 
Hadron' stands for the hadronic phase, \lq QGP' for the quark-gluon 
plasma phase and \lq CSC' for several phases that include the color 
super conducting phases. 
\item{Fig. 2.} ${\cal V} (M)$ in the case of $C_1 > 0$ against $M$. The 
curve with \lq\lq 0 '' is the ${\cal V} (M)$ with Eq. (\ref{soll}) 
for ${\cal B} (M)$. The curves with \lq\lq + '' and \lq\lq - '' is 
${\cal V} (M)$ in Eq. (\ref{redeqa}) with $|C1| C_3 / C_2^2 = + 2$ 
and $- 2$, respectively. 
\item{Fig. 3.} ${\cal V} (M)$ in the case of $C_1 < 0$. (a) The curves 
(s1), (s2), (s3) and (s4) are ${\cal V} (M)$ with, in respective 
order, the solutions (s1), (s2), (s3) and (s4) for ${\cal B} (M)$. 
(b) Actual effective potential. (c) The curve with \lq\lq 0 '' is as 
in Fig. (b) and the curves with \lq\lq + '' and \lq\lq - '' are 
${\cal V} (M)$ in Eq. (\ref{redeqa}) with $|C1| C_3 / C_2^2 = + 2$ 
and $- 2$, respectively. 
\item{Fig. 4.} $\alpha_s (\mu_B)$ against $600$ MeV $\leq \mu_B \leq 700$ 
MeV. 
\item{Fig. 5.} (a) The ratio $\rho / \rho_0$ against $0.6 \; \mbox{GeV} 
\leq \mu_B \leq 1.6$ GeV. (b) Same as (a) for $0.6 \mbox{GeV} \leq 
\mu_B \leq 4$ GeV. (c) The ratio $\rho_i / \rho$ ($i = u, d, s, c$) 
against $\mu_B$. 
\item{Fig. 6.} $C_1 (\mu_B)$ against $600$ MeV $\leq \mu_B \leq 700$ MeV. 
$d_1 = 1, 50$ and $2 \times 10^4$ for $u$, $d$ and $s$ quarks.
\item{Fig. 7.} $|\langle \mu' \rangle| = \sqrt{4 |C_1|^3 / (27 C_2)}$ 
against $600$ MeV $\leq \mu_B \leq 700$ MeV. $d_2 = 10^{-4}, 2 
\times 10^{-3}$ and $0.1$ for $u$, $d$ and $s$ quarks.
\item{Fig. 8.} (a) Relation between $\mu_L$ and $\mu_B$. (b) The ratio 
$\rho_i / \rho$ ($i = e, \mu$) against $\mu_L$. 
\label{f5} 
\end{description} 
\end{document}